\documentclass[aps,prl,twocolumn,bibnotes,nobibnotes,groupedaddress]{revtex4-1}

\usepackage{amsmath,mathrsfs,latexsym,amssymb,graphicx,xcolor,float,setspace,
graphics,subfigure}

\usepackage{hyperref}% add hypertext capabilities
\hypersetup{
colorlinks=true,%
linkcolor=black,%
linkbordercolor=0 0 0,
urlcolor=black,%
citecolor=blue,%
pdftex=true,%
bookmarks=blue}

\usepackage{lineno}

\begin{document}

%\preprint{}

\title{Lattice Induced Transparency in Metasurfaces }


\author{Manukumara Manjappa}
%\thanks{}
\affiliation{Center for Disruptive Photonic Technologies, Division of Physics and Applied Physics, School of Physical
and Mathematical Sciences, Nanyang Technological University, 21 Nanyang Link, Singapore 637371.
}

\author{Yogesh Kumar Srivastava}
%\thanks{}
\affiliation{Center for Disruptive Photonic Technologies, Division of Physics and Applied Physics, School of Physical
and Mathematical Sciences, Nanyang Technological University, 21 Nanyang Link, Singapore 637371.
}

\author{Ranjan Singh}
\email[]{ranjans@ntu.edu.sg}
%\thanks{}
\affiliation{Center for Disruptive Photonic Technologies, Division of Physics and Applied Physics, School of Physical
and Mathematical Sciences, Nanyang Technological University, 21 Nanyang Link, Singapore 637371.
}


%\date{\today}

\begin{abstract}
Lattice modes are intrinsic to the periodic structures and their occurrence can be easily tuned and controlled by changing the lattice constant of the structural array. Previous studies have revealed excitation of sharp absorption resonances due to lattice mode coupling with the plasmonic resonances. Here, we report the first experimental observation of a lattice induced transparency (LIT) by coupling the first order lattice mode (FOLM) to the structural resonance of a metamaterial resonator at terahertz frequencies. The observed sharp transparency is a result of the destructive interference between the bright mode and the FOLM mediated dark mode. As the FOLM is swept across the metamaterial resonance, the transparency band undergoes large change in its bandwidth and resonance position. Besides controlling the transparency behaviour, LIT also shows a huge enhancement in the \textit{Q}-factor and record high group delay of 28 \textit{ps}, which could be pivotal in ultrasensitive sensing and slow light device applications.   
\end{abstract}


\maketitle


\section{Introduction}
Metamaterials (MMs)\citep{1,2} are arguably one of the simplest materials to control and engineer the resonance properties of the light-matter interaction by altering the geometry of their structures. Perfect lens\citep{3,4}, Negative refraction\citep{5}, invisibility cloak\citep{6}, phase engineering and many more intriguing phenomenon have been realized by exploiting the impressive properties offered by the MMs. In recent years, engineering resonant transmission properties in the terahertz band of the electromagnetic spectrum using MM structures has gained tremendous interest for achieving anomalous behaviour in the phase of the propagating field. These devices are of immense application in broadband telecommunication networks as efficient modulators and phase retarders at terahertz frequencies. One way of tailoring the phase anomalies using metamaterials is by inducing a sharp transparency through intra unit cell coupling of bright-dark resonances via Fano-type of interference\citep{7,8,9,10,11}. In this work, we show a powerful and an efficient way of tailoring the transmission and phase anomalies by diffraction coupling the plasmonic resonances. This coupling is mediated and tailored by the lattice mode, which can be adequately controlled by varying the lattice constant of the metamaterial structure.  

\begin{figure}[h]
\begin{center}
  \includegraphics[width=8cm]{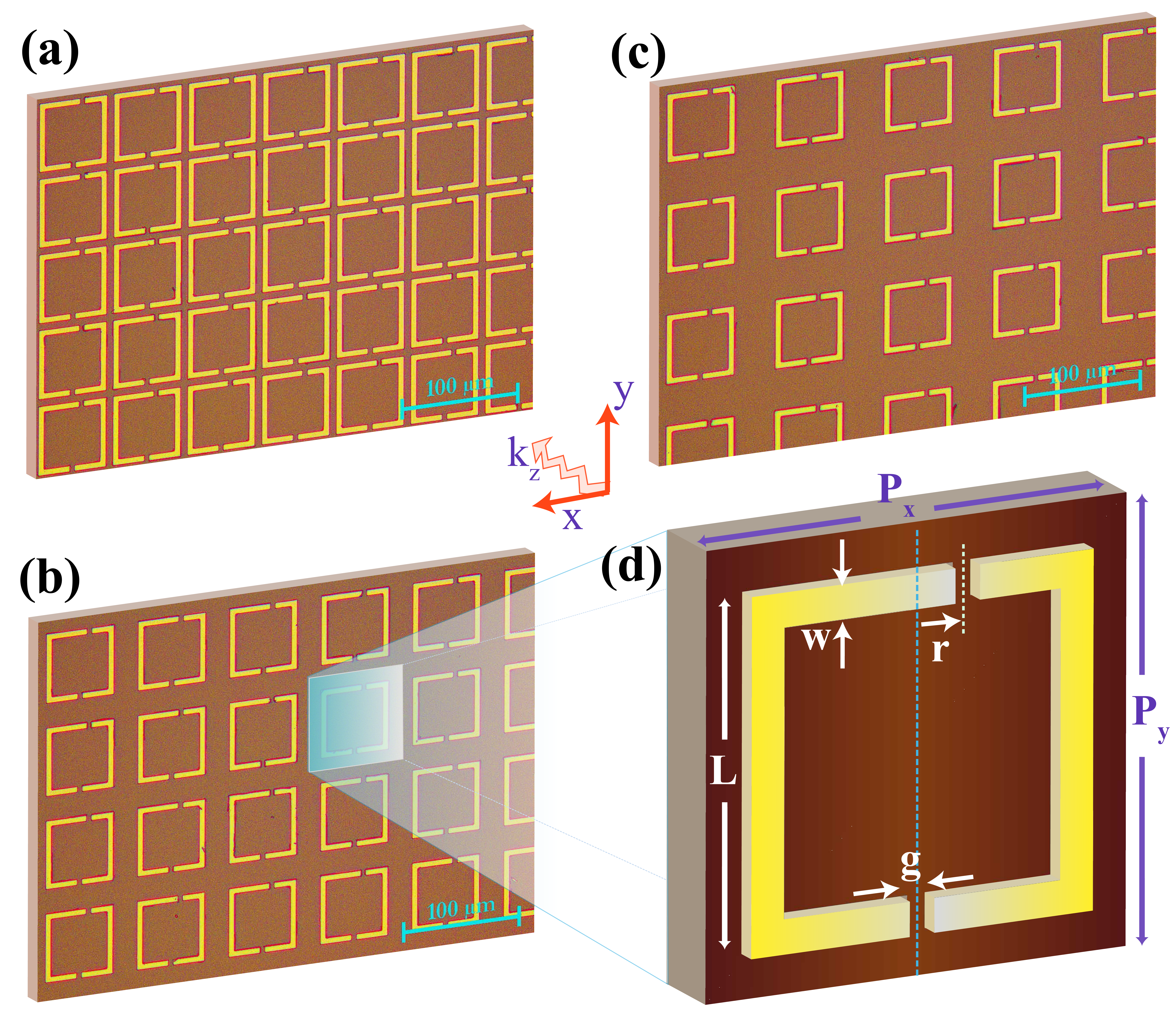}
  \caption{\textbf{(a)}, \textbf{(b)} and \textbf{(c)} are the Optical Microscopy (OM) images of the metamaterial samples with lattice constant/periodicity $P_{x} = P_{y} = P = $ 65 $\mu$m, 80 $\mu$m and 95 $\mu$m, respectively. Subfigure \textbf{(d)} depicts the unit cell dimensions of the metamaterial sample. Length, \textbf{L}: 60 $\mu$m; asymmetry, \textbf{r}: 8 $\mu$m; gap, \textbf{g}: 3 $\mu$m; and width, \textbf{w}: 6 $\mu$m. } \label{fig1}
\end{center}
\end{figure}

Lattice mode resonances often appear due to the discontinuity in the dispersion curves at the ambient medium-substrate interfaces at Rayleigh cut-off wavelengths\citep{12,13} of the incident field. These are also referred to as diffractive modes or Wood anomalies\citep{14}, whose spectral position depends on the periodicity of the unit cell and the incident angle of the excitation field. The anomalies appearing due to diffraction are intrinsic to subwavelength periodic structures such as metamaterials and despite their attractive features like extremely narrow resonances and broadband tunability that depends on the lattice geometries, they have not been much explored. In the past there has been a handful number of works that demonstrated the implications of these lattice modes in the periodic structures. Coupling the plasmonic resonances to the lattice modes has yielded resonant suppression in the light extinction\citep{15} and observed extremely high \textit{Q} resonances at near infrared\citep{16,17,18} and far infrared frequencies\citep{19} and these observations have not shown any signature that alters the nature of the plasmonic resonance itself. Recently, S. D. Jenkins et al.\citep{20} proposed the excitation of the sharp transparency in the reconfigurable MM array by actively coupling their plasmonic resonances to the lattice excitation. In this work, we experimentally demonstrate for the first time a lattice induced transparency (LIT) by coupling the first order lattice mode (FOLM) to the MM structural mode of the periodically arranged terahertz asymmetric split ring resonators (TASRs) MM structure. Our results show large enhancement in the \textit{Q}-factor and group delay for the transmission resonance in the structure, where the FOLM is exactly matched with the MM resonance.     

\section{Results and Discussion}
To investigate the influence of the lattice mode on the MM resonances, we considered MM samples consisting of TASR with varying square lattice constants as shown in Fig. 1. Samples were fabricated by using the conventional photolithography technique, where 200 nm thick aluminium metal structures were deposited on a 500 $\mu$m thick high resistivity silicon wafer using thermal evaporation technique followed by the lift-off process.  Fig. 1(d) shows the unit cell dimensions of the TASR MM samples with varying lattice constants/periodicities, \textit{P} = 65 $\mu$m, 80 $\mu$m and 95 $\mu$m. These MM samples are designed in such a way that the nature and position of MM resonances are unaltered by keeping the geometry of the unit cell same for the all three samples, whereas the frequency of the lattice mode is continuously tuned by varying the lattice constant along \textit{x} and \textit{y} axis of the sample. The diffracted frequency for the lattice is determined by the following simplified expression for a periodic square lattice with normal beam incidence,  
\begin{equation}
f_{diff} = \frac{c}{n P} \sqrt{i^{2}+j^{2}} \label{eqn1}
\end{equation}
where, \textit{c} is the speed of light in vacuum, \textit{n} is the refractive index of the substrate, \textit{P} is the lattice constant of the structure and \textit{(i,j)} are the pair of indices defining the order of the lattice mode. For the first order lattice mode (FOLM) \textit{(0,1)}, the above expression reduces to $f_{diff} = \frac{c}{n P}$ , where the frequency of the lattice mode varies inversely with the lattice constant (\textit{P}). \\

\begin{figure}[h]
\begin{center}
  \includegraphics[width=8cm]{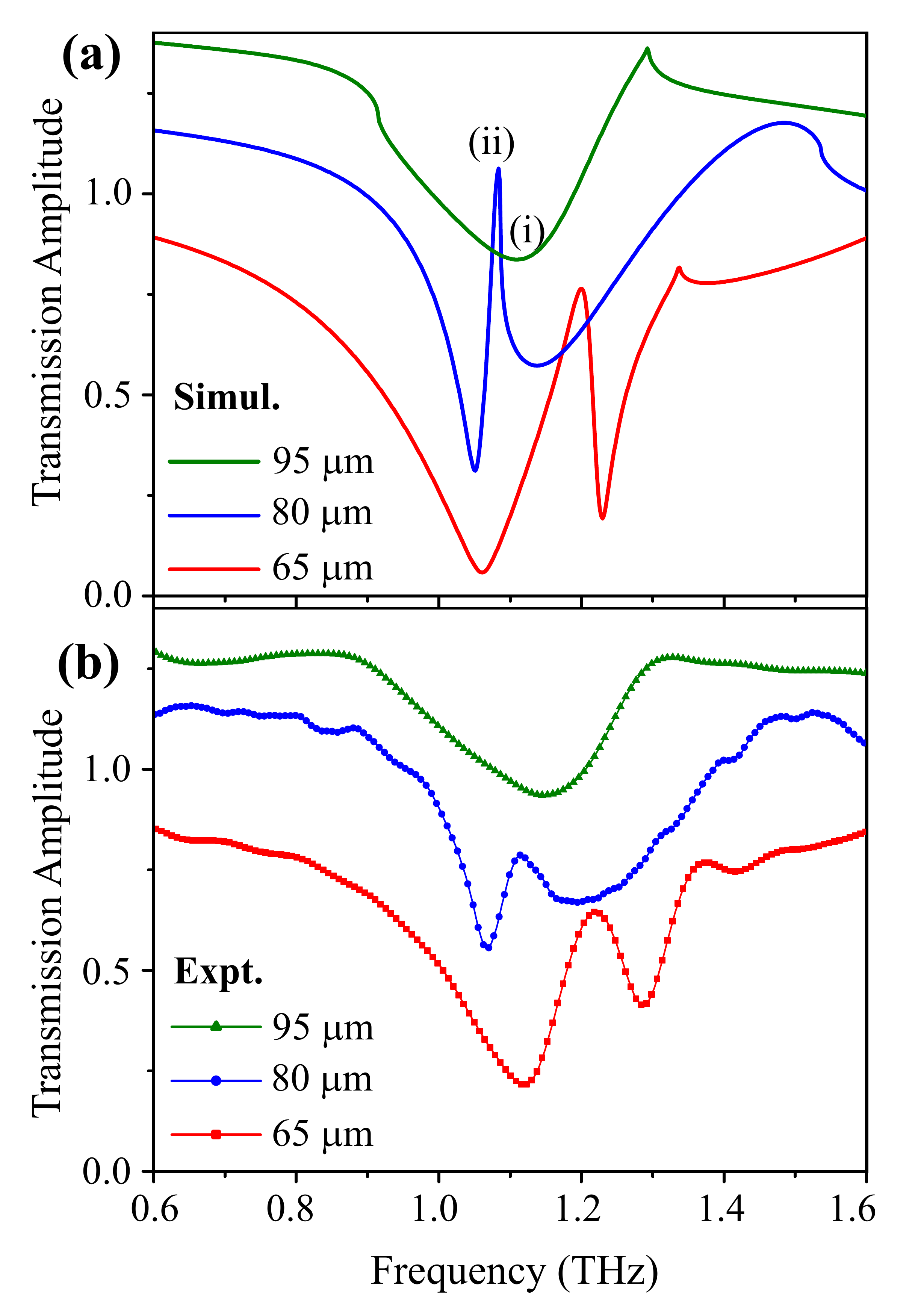}
\end{center}
\caption{\textbf{(a)}: Numerically simulated curves  depicting the modulation of transmission window for various lattice constants (\textit{P} = 65 $\mu$m, 80 $\mu$m and 95 $\mu$m) of the metamaterial structure. \textbf{(b)}:  Experimental curves justifying the numerically predicted results. The curves are drawn with vertical offset for better visibility. } \label{fig2}
\end{figure} 

In this work, we numerically as well as experimentally investigate the effect of FOLM on the dipolar resonance appearing at 1.1 THz for the incident polarization \textbf{$E_{x}$}. Effect of the lattice mode on the MM resonance was primarily investigated by rigorous numerical simulations using the finite integral techniques method offered by the CST Microwave Studio Frequency Solver. Here, it is worth mentioning that in the simulations, where the resonances possess high \textit{Q}-factors, choice of the material plays a crucial role\citep{21}. Materials for the numerical simulation are chosen to be aluminum metal with DC conductivity ($\sigma_{DC}$ = 35.6 MS/m) as the TASR metallic resonators and silicon with dielectric constant $\epsilon$ = 11.9 used as the transparent substrate at terahertz frequencies. The lattice constant (\textit{P}) for the MM unit cell is chosen in such a way that for \textit{P} = 80 $\mu$m, the FOLM is precisely matched to the TASR resonance at 1.1 THz, whereas for \textit{P} = 65 $\mu$m and \textit{P} = 95 $\mu$m, FOLM is towards the blue and the red side of the TASR resonance, respectively. Green curve in Fig. 2 (a) depicts the simulated transmission response for the MM sample with lattice constant, \textit{P} = 95 $\mu$m for the \textbf{$E_{x}$} excitation. Despite possessing the structural asymmetry, no mode splitting due to bright-dark coupling is observed as reported by some of the previous works\citep{22,23,24}. For \textbf{$E_{x}$} excitation, we see that the dark mode due to structural asymmetry remains unexcited in this system, showing the symmetric dipolar currents running parallel to the polarization of the excitation field, \textbf{$E_{x}$} (see Fig. 3(i)). The observed asymmetry of this absorption dip is due to the weak coupling of the MM structural resonance with the FOLM appearing at 0.9 THz. The other sharp feature observed at about 1.3 THz corresponds to the higher order \textit{(1,1)} lattice mode.  \\ 

As the periodicity of the structure is decreased to 80 $\mu$m, we notice a strong coupling of the MM eigen resonance to the FOLM, which results in a sharp transparency at the center of the TASR MM resonance, as shown by the blue curve in Fig. 2(a). When the \textbf{$E_{x}$} polarized light is incident on the sample, broad dipolar resonance centered at 1.1 THz interacts resonantly with weakly oscillating FOLM that propagates on the surface of the substrate. Lattice mode whose energy gets trapped within the MM array couples to the other surface modes rather than to the free space and mediates the excitation of the otherwise unexcited dark mode within the TASR MM system. This interaction between the broad MM resonance and the weakly excited dark resonance that is mediated by the lattice mode undergo Fano-type of interference to give rise to a sharp transmission window.  As the lattice constant is decreased to 65 $\mu$m, the FOLM is blue shifted with respect to the MM resonance, which results in the blue shift of the observed transmission peak, as shown by the red curve in Fig. 2(a). Observed frequency shift of the transparency peak with lattice constant of the unit cell affirms our claim that the resulted transparency is mediated by the lattice mode propagating on the surface of the substrate. Our simulation results are experimentally verified by performing the 200 \textit{ps} long scan THz-Time Domain Spectroscopy (THz-TDS) measurements on the fabricated TASR MM samples. The post processed transmission spectra of the measured THz time domain pulses are shown in Fig. 2(b), which matches well with the simulation results.   \\ 

\begin{figure}[h]
\begin{center}
  \includegraphics[width=8cm]{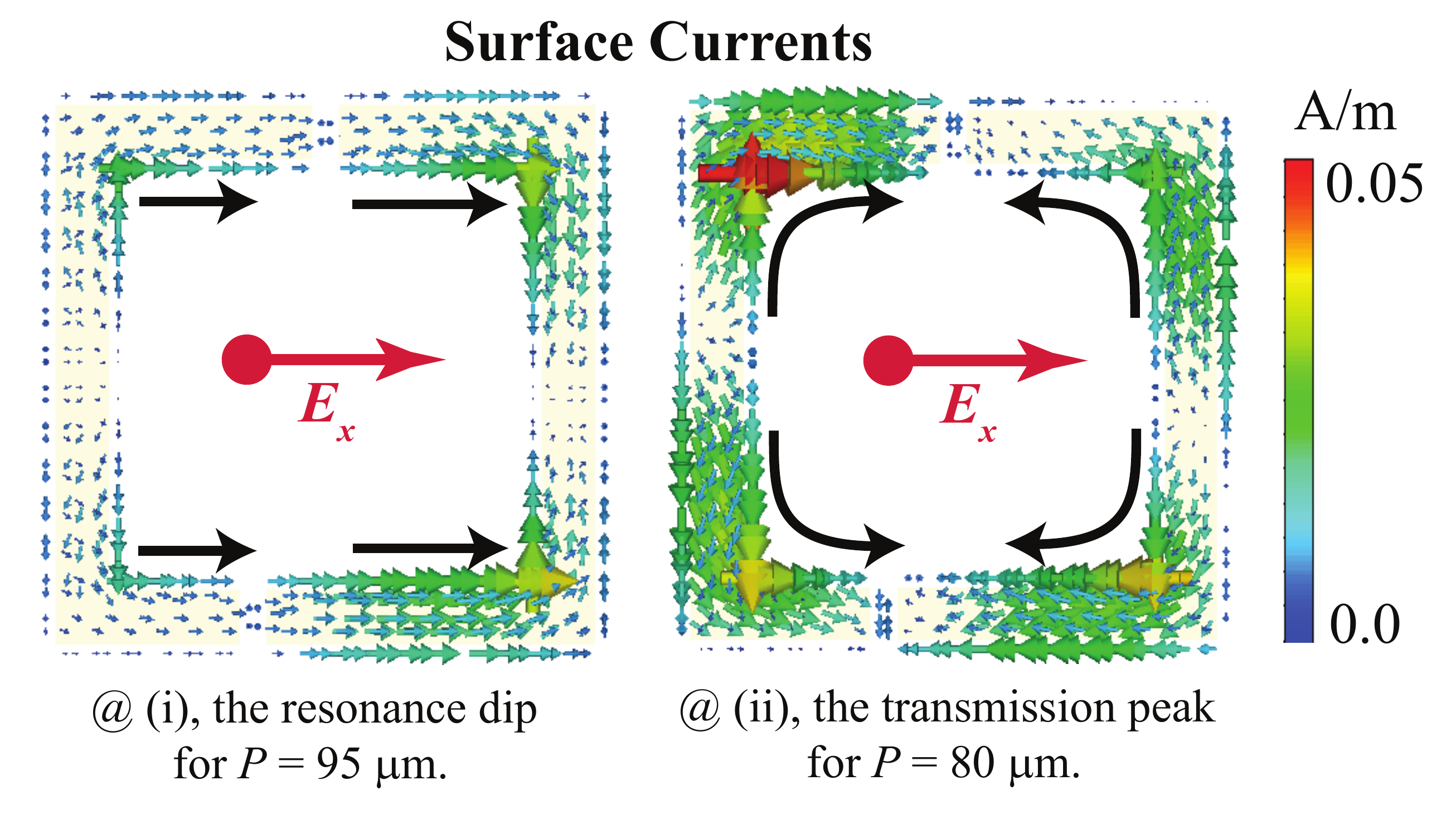}
\caption{Surface current distribution at resonance dip (i) for \textit{P} = 95 $\mu$m and at the transparency peak (ii) for \textit{P} = 80 $\mu$m for the curves shown in Fig. 2(a). } \label{fig3}
\end{center}
\end{figure}

The underlying phenomenon behind the LIT is further understood by studying the surface current distribution at the resonance dip (i) for \textit{P} = 95 $\mu$m and at the transparency peak (ii) for \textit{P} = 80 $\mu$m as shown in Fig. 3. For the MM eigen resonance that is not matched to the FOLM (for \textit{P} = 95 $\mu$m) results in a dipolar type of surface currents, where the currents on the resonator arms run parallel to the electric component of the excitation field, \textbf{$E_{x}$}. When the MM dipolar resonance is matched to the FOLM (for \textit{P} = 80 $\mu$m), whose energy is propagated on the surface of the substrate interacts with the MM dipole resonance to induce the antiparallel currents along the opposite arms of the resonator. These opposing currents are quadrupolar in nature and destructively interfere to result in the cancellation of the fields at the resonance leading to a sharp transmission peak and steep dispersion within the system. Thus the nature of the MM resonance is modified by the influence of the lattice mode, which alters the dipolar type of resonance into a sharp transmission feature showing the quadrupolar nature of the surface currents.\\

\begin{figure}[h]
\begin{center}
  \includegraphics[width=8cm]{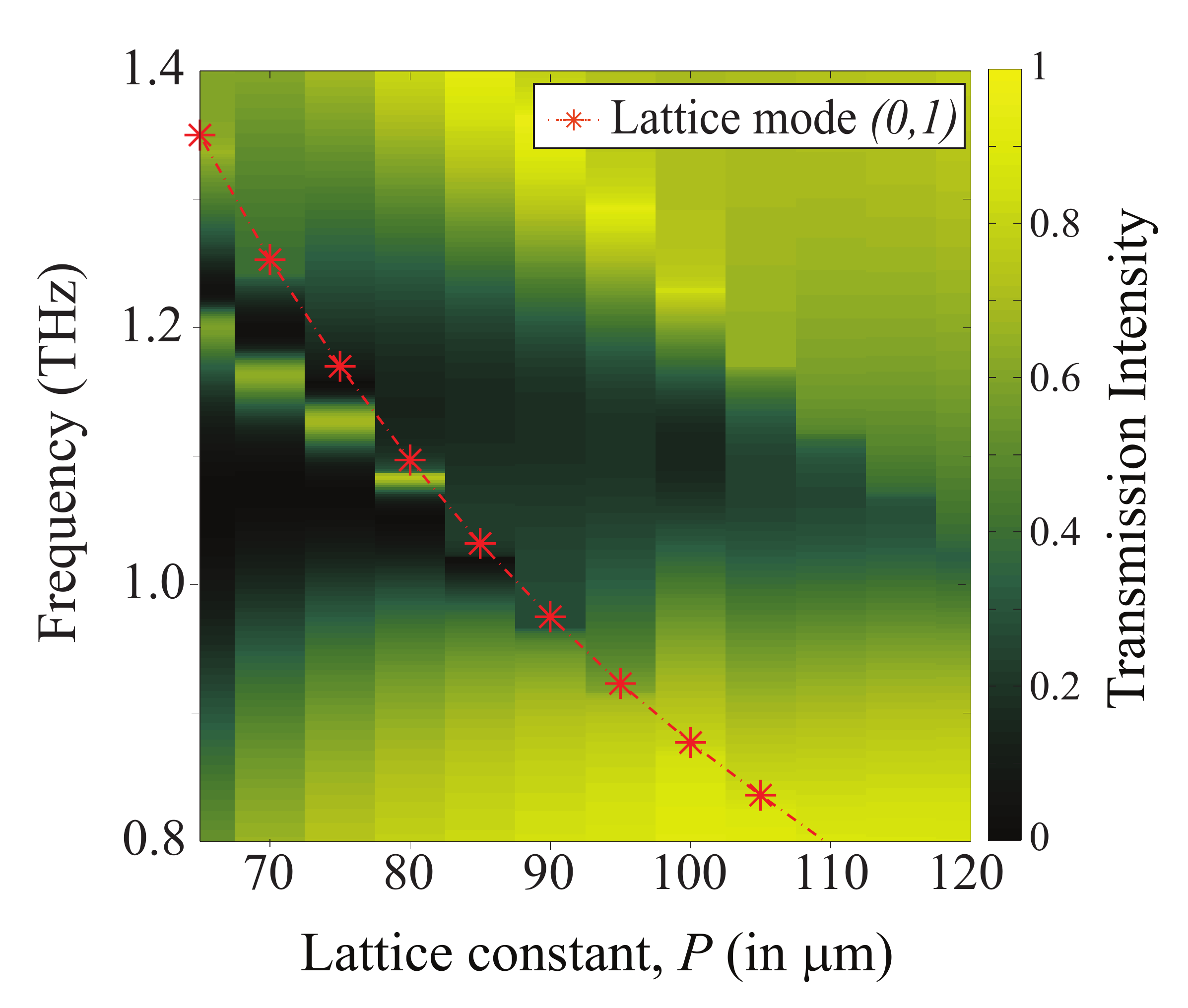}
\end{center}
\caption{Contour plot showing the variation in the transmission intensity of the LIT resonance as a function of frequency versus the lattice constant. The red dash-dotted curve signifies the resonance variation of first order lattice mode (FOLM) \textit{(0,1)} with varying lattice constants (\textit{P}). There is a clear indication that the change in the position of the transparency peak follows the variation of FOLM. } \label{fig4}
\end{figure} 

Fig. 4 represents the contour diagram showing the intensity variation of the transmitted field with respect to the frequency of the incident beam and the lattice constants of the MM structure. Variation in the frequency of the FOLM with respect to the lattice constant is also plotted and is shown by the red dot-dashed lines in Fig. 4. As observed, shift in the frequency of the transparency peak follows the frequency shift of the FOLM. As the FOLM approaches the TASR MM resonance, we observe a gradual narrowing of the resulting transmission peak. This enhanced sharpness in the transmission window can be attributed to the increased diffraction effects near the resonance that traps the energy in the MM array and suppresses the radiation propagating into the far-field. The extracted \textit{Q}-factors of the LIT band from the simulated transmission curves are 25, 30, 35 and 91 for \textit{P} = 65 $\mu$m, 70 $\mu$m, 75 $\mu$m and 80 $\mu$m, respectively. For the observed transparency peaks, we extract the Figure of Merit (FoM), which quantifies the strength of the resonance in terms of product of the \textit{Q}-factor and intensity of the transparency peak (FoM = $Q \times \Delta I$). FoMs for the transparency peak for the lattice constants, \textit{P} = 65 $\mu$m, 70 $\mu$m, 75 $\mu$m and 80 $\mu$m are 13.7, 18.4, 23.6 and 55, respectively. For \textit{P} = 80 $\mu$m, where the FOLM is exactly matched to the resonance of the TASR mode, we observe a steep rise in the \textit{Q}-factor (91) and FoM (55) for the transparency that is higher than the values so far reported for these structures\citep{22}.    \\

\begin{figure}[h]
\begin{center}
  \includegraphics[width=8cm]{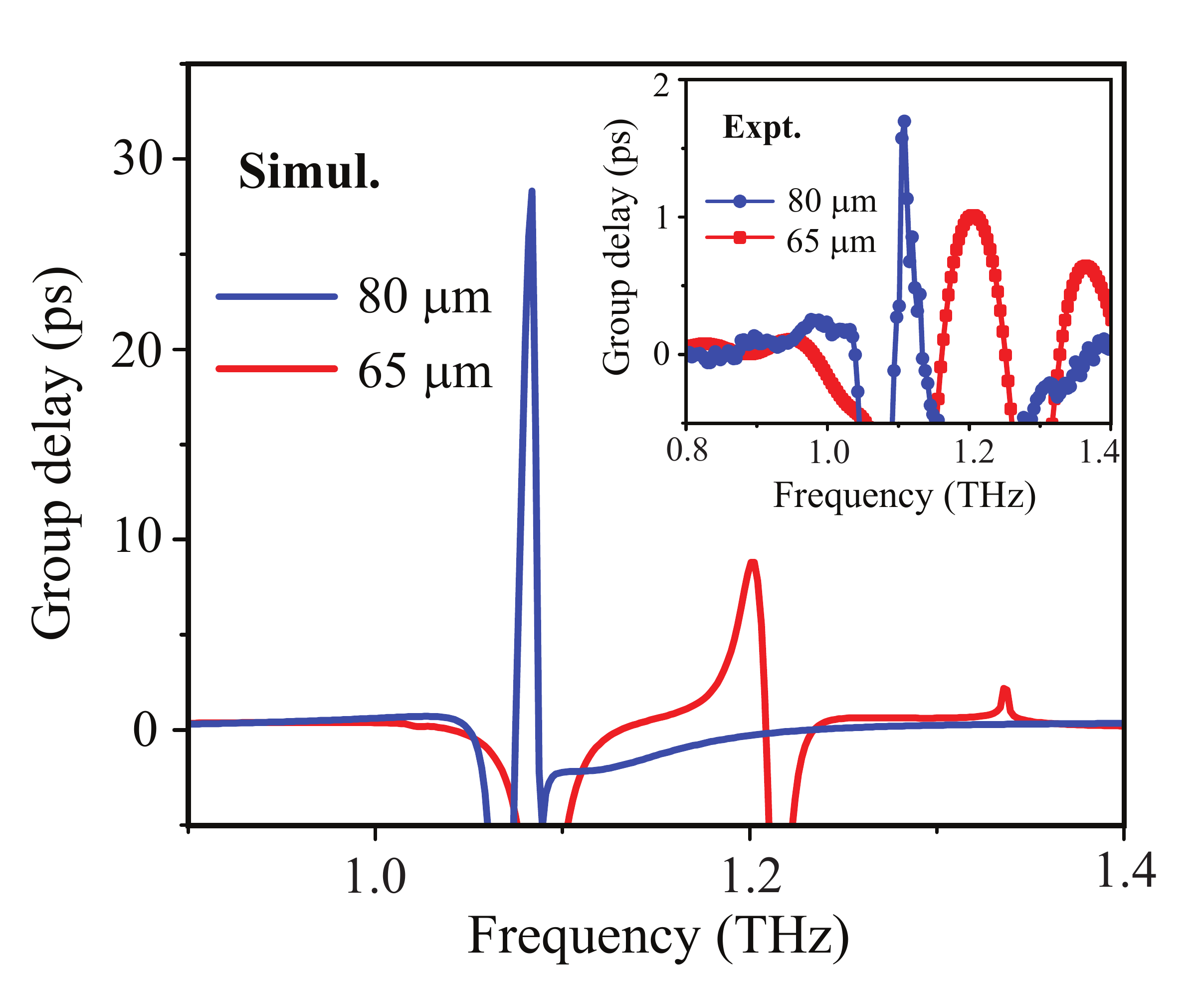}
\caption{Numerically obtained group delay plots for lattice constants, \textit{P} = 65$\mu$m, 80$\mu$m. Inset figure represents the group delay curve extracted from the measured data.  }\label{fig5}
\end{center}
\end{figure}

The improved strength in the transparency peak observed due to diffraction effects induces an anomalous resonant dispersion that enhances the slow-light behaviour within the system. The slow-light nature of the pulse is quantified in terms of the group delay values extracted from the phase of the pulse propagating through the medium by using the expression, $t_{g}=\frac{-d\phi}{d\omega}$, where $\phi$ is the phase variation of the pulse propagating through the medium and $\omega$ is the angular frequency of the incident pulse. As illustrated in Fig. 5, our simulation results estimate as high as 28 \textit{ps} of delay in the group velocity of the terahertz pulse through the MM sample with lattice constant, \textit{P} = 80 $\mu$m, whereas for \textit{P} = 65 $\mu$m the group delay is around 10 \textit{ps}.  A sharp increase in the group delay value for \textit{P} = 80 $\mu$m is attributed to the strong interaction of the MM resonance with the FOLM, that slows down the propagation of light quite drastically. Coupling the plasmonic resonances with the nearby lattice mode gives a great advantage of tailoring the group delay of the incident terahertz pulse near the resonance with a remarkable enhancement in the slow light behaviour within the system. These lattice coupled systems can be of great deal of interest in many real world applications such as in fast growing broadband communication technologies as optical routers and phase retarders. The group delay values shown in the inset of Fig. 5 are extracted from the measured data and shows similar trend as observed in the simulations. The observed discrepancies in the absolute values of the measured and simulated group delay values can be attributed to the limited scan length offered by the THz-TDS measurement setup to experimentally capture the sharp transmission resonance appearing due to the diffraction effects. Capturing the higher group delay values are always challenging in the laboratory experiments as the measurements can be altered by the undesired substrate effects, fabrication errors, and errors due to incident polarization and the incident angle of the excitation field.  

\section{Conclusion}
In conclusion, we have numerically and experimentally demonstrated the lattice induced transparency in metasurfaces by radiatively coupling the metamaterial resonances with the fundamental diffractive lattice mode (FOLM). The nature of the LIT transmission band was modulated by sweeping the lattice mode across the metamaterial eigen resonance by changing the periodicity of the metamaterial sample. Besides the observed transparency window, our results demonstrate a very sharp transmission feature with significantly enhanced group delay for the incident terahertz pulse. In addition to the enhanced slow-light effects, lattice mediated transparency can be pivotal in achieving ultra-high \textit{Q} transmission resonances for sensing applications, energy harvesting devices and in realizing sharp delay lines and optical phase shifter in broadband telecom applications.          

\section{Methods}

\textbf{Numerical simulations}: Numerical calculations were performed using Finite Integral Techniques offered by the commercially available simulation software; Computer Software Technology (CST) Microwave studio Frequency Solver. The lossy aluminium metal with DC conductivity, $\sigma_{DC} = 3.56 \times 10^{7}$ S/m was modeled as the terahertz asymmetric ring resonator (TASR) and the loss less silicon with permittivity, $\epsilon$ = 11.9 was chosen as the transparent substrate. The terahertz wave with the electric component of the field oscillating perpendicular to the TASR   is normally incident on the designed metamaterial (MM) surface. In the simulation, a single unit cell of the TASR MM structures was simulated using the tetrahedral mesh geometry with unit cell boundary conditions employed in the planar directions orthogonal to the incident field. In the simulations, contribution from all the modes occurring within the 0 to 2 THz bandwidth was considered. The scattering (\textit{S}) parameter, $S_{21}$ that represents the transmission spectra is shown in the Fig. 2(a). The field monitors were used to simulate the surface currents at the resonance frequencies of the absorption and the transmission peak of the spectrum.\\

\textbf{Fabrication process}: The TASR MM samples were fabricated using the conventional photolithography technique. A positive photoresist was coated on a 500 $\mu$m thick double side polished high resistivity silicon ($\rho$ $>$ 5000 $\Omega$-cm) wafer using the spin coating technique followed by the prebaking process at 100$^{\circ}$ C for 1 minute. After prebaking, the photoresist was covered with the positive mask and is exposed to the UV light. After, the UV exposed wafer was soaked in developer solution to remove the UV exposed area of the photoresist. Later, 200 nm thick aluminum metal (Al) was deposited on the developed wafer using the thermal evaporation process. To have the desired TASR pattern of the aluminum metal, the Al deposited wafer was put in acetone for few hours to liftoff the undesired photoresist from the surface of the wafer. \\

\textbf{Transmission measurements} : Terahertz time domain spectroscopy (THZ-TDS) measurement technique is used to characterise the fabricated TSAR MM samples for the normal incidence of the THz pulse. The incident THz pulse has a spot size of 3 mm and is polarized perpendicular (\textbf{$E_{x}$}) to the TASR gaps. The measurement was performed in the time domain by using the standard GaAs photoconductive antenna based terahertz system that offers high signal to noise ratio (10000:1). To perform the long scan measurements ($\sim$ 200 \textit{ps}), we optically glued MM sample to a 15 mm thick double side polished silicon wafer of high resistivity ($\rho$ $>$ 5000 $\Omega$-cm) using acetone. This helps us in delaying the reflection pulse appearing due to fabry-perot reflection within the substrate. The performed long scan measurement improves the frequency resolution of our detection to 5 GHz. The measured time pulses are fast Fourier transformed (FFT) in the post processing steps to extract the transmission spectrum and is shown in the Fig. 2(b).

\end{document}